\documentclass[preprint, amsmath,amssymb,aps,showpacs]{revtex4}
\usepackage[english]{babel}
\usepackage{graphicx}
\usepackage{graphics}
\usepackage{amsmath}
\usepackage{dcolumn}
\usepackage{amssymb}
\usepackage{bm}
\usepackage[latin1]{inputenc}

\begin{document}
\title{Thermal Casimir effect for neutrino and electromagnetic fields 
in closed Friedmann cosmological model}
\author{V. B. Bezerra,$^1$ V. M. Mostepanenko,$^{1,2}$ 
H. F. Mota,$^1$ and C. Romero$^1$}
\affiliation{${}^1$Department of Physics, Federal University of 
Paraíba, C.P. 5008, 58051-970, Jo\~ao Pessoa,
PB, Brazil \\ ${}^2$Noncommercial Partnership ``Scientific Instruments'', 
Tverskaya Street 11, Moscow,
103905, Russia}

\begin{abstract}
We calculate the total internal energy, total energy density and pressure,
 and the free energy
for the neutrino and electromagnetic fields in Einstein and closed 
Friedmann cosmological models.
The Casimir contributions to all these quantities are separated. 
The asymptotic expressions for
both the total internal energy and free energy, and for the Casimir 
contributions to them are
found in the limiting cases of low and high temperatures. It is shown that the neutrino field
does not possess a classical limit at high temperature. As for the electromagnetic field,
 we demonstrate that the total internal energy has the classical contribution and the Casimir
 internal energy goes to the classical limit at high temperature. The respective Casimir free
 energy contains both linear and logarithmic terms with respect to the temperature. The total
 and Casimir entropies for the neutrino and electromagnetic fields at low temperature are also
 calculated and shown to be in agreement with the Nernst heat theorem.
\end{abstract}
\pacs{04.62.+v, 44.40.+a, 98.80.-k}
\maketitle

\section{Introduction}

The Casimir effect \cite{01} manifests itself as some vacuum polarization energy 
and force due to
material boundaries or nontrivial topology of space. This effect is presently 
well known owing to
its multidisciplinary applications ranging from condensed matter physics and nano\-technology to
quantum field theory, gravitation and cosmology (see monographs \cite{02,03,04,05}). During the
last decade, numerous experiments on measuring the Casimir force have been performed as reported
in reviews \cite{06,07,08}. Their results were used to obtain stronger constraints on 
the parameters
of the Yukawa-type corrections to Newtonian gravitational law predicted in many 
extensions of the
standard model (see the most recent results in Refs. \cite{09,10,11,12,13}). 
Keeping in mind that
 measurements of the Casimir force are usually performed at room temperature, 
the problem of thermal
 Casimir effect became topical. In this regard it was found that the influence 
of free charge carriers
 in the boundary bodies on the thermal correction to the Casimir force is highly nontrivial and
 leads to complicated problems in quantum statistical physics \cite{05,06,08,14}.

In cosmology, the Casimir effect arises not due to the material boundaries, but due to the
identification conditions imposed on the wave functions by a nontrivial topology of space.
Specifically, the Casimir energy density and pressure for the scalar field in the Einstein and
closed Friedmann cosmological models with a topology $S^{3}\times R^{1}$ were found in
Refs. \cite{15,16} and in Ref. \cite{17} in the massless and massive case, respectively.
The Casimir density and pressure for the neutrino and electromagnetic fields in the Einstein
cosmological model were derived in Ref. \cite{18}. In succeeding years the Casimir effect was
investigated in many topologically nontrivial spaces \cite{19,20,21,22}. This resulted in the
development of new methods in mathematical physics and found prospective application in
multidimensional cosmology \cite{23,24} and in the problem of dark energy \cite{25}.

A considerable number of works was devoted to the thermal Casimir effect in cosmological
models when not only zero-point photons, but thermal radiation as well was taken into account.
Specifically, thermal Green's functions in Rindler, de Sitter and Schwarzschild spaces were
considered in Ref. \cite{26}. In Refs. \cite{27,28} the thermal stress-energy tensor of the
scalar field in the Einstein cosmological model was investigated and the asymptotic behaviors
at low and high temperature were found. Similar problem for a cosmological model with 
a $3$-torus
topology was solved in Ref. \cite{29}. In Ref. \cite{30} the results of Refs. \cite{27,28} were
reconsidered using another approach. It was shown that the total thermal stress-energy tensor of
the scalar field in Einstein model contains the Casimir contribution which has the 
same asymptotic
 behavior at low and high temperature as in the case of two parallel ideal-metal plates.
 The total thermal stress-energy tensors of the neutrino and electromagnetic fields in Einstein
 cosmological model were considered in Ref. \cite{31}. The expressions obtained were
 used \cite{31,32} to determine the back reaction of the total thermal stress-energy tensor
 on the space-time by solving the Einstein equations, where this tensor would play the 
role of a source.

It is common knowledge that at the very early stages the evolution of our universe 
has been going on at very high temperatures. Because of this, the studies of
thermal Casimir effect in cosmological models mentioned above are of great physical
significance. Keeping in mind, however, that massless fields of zero spin are
not observed in nature, main attention in this respect should be paid to neutrino
and electromagnetic fields. Considering that at large scales our universe is
spatially homogeneous,
 here we derive exact
expressions for the total thermal stress-energy tensor for these fields
in Einstein and closed Friedmann cosmological models. 
We also find the total
and Casimir free energy and the Casimir contributions to the stress-energy tensor 
(the latter were
 previously known only at zero temperature). We find the asymptotic behaviors of the obtained
 expressions in the limiting cases of low and high temperatures. For the neutrino case 
our results
 for the total energy and pressure are the same as in Ref. \cite{31}. For the 
electromagnetic case,
  however, there is a disagreement between our thermal stress-energy tensor and that obtained in
  Ref. \cite{31}. We explain this disagreement by the omission in \cite{31} of a nontrivial
  contribution arising from the zeroth mode in the Poisson (or, equivalently, Abel-Plana)
  summation formula. The expression for the Casimir free energy and Casimir internal energy for
  the electromagnetic field obtained by us are in direct analogy with the familiar case of the
  thermal Casimir effect in an ideal metal spherical shell. For both the neutrino 
and electromagnetic
fields we calculate the Casimir entropy and demonstrate that the third law of thermodynamics
(the Nernst heat theorem) is followed.

The structure of the paper is the following. In Sec. II we present general expressions for the
free energy, internal energy and stress-energy tensor of the neutrino and electromagnetic fields
in the Einstein cosmological model. Special attention is paid to the renormalization 
procedure and
to the definitions of the Casimir contributions to the considered quantities at 
nonzero temperature.
Section III is devoted to the derivation of the total and Casimir free energy and 
internal energy
for the neutrino field. The total and Casimir pressures are also considered. The asymptotic
expressions of the obtained expressions at low and high temperatures are obtained and the
validity of the Nernst heat theorem is verified. Similar results for the thermal Casimir effect
in the electromagnetic case are presented in Sec. IV. Here, we concentrate on delicate points
in the applicability of summation formulas and analogies with the case of electromagnetic field
inside an ideal metal spherical shell. Our conclusions and discussions are presented in Sec. V.

For convenience in comparisons with the classical limit (the case of high 
temperature \cite{33}),
below we retain the Planck constant $\hbar$, the velocity of light $c$, 
and the Boltzmann constant
$k_B$ in all equations.

\section{General expressions for the total and Casimir free ener\-gy and internal energy
for neutrino and electromagnetic fields }

We consider the four component neutrino field and the electromagnetic field in 
thermal equilibrium
at some temperature $T$ in the static Einstein cosmological model with a 
topology $S^{3}\times R^{1}$.
The metric of this model is given by
\begin{equation}
ds^2=c^2d\tau^2-a^2[dr^2+\sin^2r(d\theta^2+\sin^2\theta d\varphi^2)],
\label{eq0}
\end{equation}
\noindent
where $r,\,\theta,\,\varphi$ are dimensionless coordinates on a three-space
of constant curvature +1 and $\tau$ is the proper synchronous time.
This is a spatially homogeneous model with a finite spatial 
volume $V=2\pi^{2} a^{3}$, where 
$a=$const is the scale factor.
 All the results obtained below are applicable also in the conformally 
static closed Friedmann
cosmological model with the scale factor depending on time (the concept 
of thermal equilibrium in
nonstationary situations is discussed in Ref. \cite{34}). In this case, 
however, there are additional
contributions to the stress-energy tensor due to conformal anomaly and 
creation of particles \cite{35, 36}.
It has long been known that the three Friedmann cosmological models
(open, quasi-Euclidean and closed) form the theoretical basis of 
modern astrophysics and cosmology. Because of this the results obtained
below are not of only academic interest but can be used in 
cosmological applications.

The free energy of the neutrino $(s=1/2)$ and electromagnetic $(s=1)$ fields is given by
\begin{eqnarray}
F^{(s)}(T)=E^{(s)}_{0}+\Delta F^{(s)}(T),
\label{0.1}
\end{eqnarray}
where $E^{(s)}_{0}$ is the zero-point energy at zero temperature and $\Delta F^{(s)}(T)$ is
the thermal correc\-tion. For the neutrino field we have \cite{18,31}
\begin{eqnarray}
E^{(1/2)}_{0}=-2\hbar\sum_{n=1}^{\infty}n(n+1)\omega_{n}^{(1/2)}, \;\;\;\;\;\
\omega_{n}^{(1/2)}=\frac{(n+\frac{1}{2})c}{a},
\label{0.2}
\end{eqnarray}
\begin{eqnarray}
& &\Delta F^{(1/2)}(T)=-4k_{B}T\sum_{n=1}^{\infty}n(n+1)\mathrm{ln}
\left[1+e^{-(\hbar\omega_{n}^{(1/2)}/k_{B}T)}\right].
\label{0.3}
\end{eqnarray}
For the electromagnetic field the respective expressions are the following \cite{18,31}:
\begin{eqnarray}
E^{(1)}_{0}=\hbar\sum_{n=1}^{\infty}(n^2-1)\omega_{n}^{(1)}, \;\;\;\;\;\ \omega_{n}^{(1)}=\frac{nc}{a},
\label{0.4}
\end{eqnarray}
\begin{eqnarray}
\Delta F^{(1)}(T)=2k_{B}T\sum_{n=1}^{\infty}(n^2-1)\mathrm{ln}\left[1-e^{-(\hbar\omega_{n}^{(1)}/k_{B}T)}\right].
\label{0.5}
\end{eqnarray}
Note that the zero-point energy $E^{(s)}_{0}$ given by Eqs. (\ref{0.2}) and (\ref{0.4}) is divergent,
whereas the thermal correction $\Delta F^{(s)}(T)$ in Eqs. (\ref{0.3}) and (\ref{0.5}) is finite.
It is notable also that the vacuum energy $E^{(1/2)}_{0}$ is negative, as it should be in the spinor case.

The renormalization of the vacuum energy $E^{(s)}_{0}$ is conventionally performed by subtracting the
terms which are obtained from Eqs. (\ref{0.2}) and (\ref{0.4}) by replacing the discrete sums with
integrations over a continuous variable \cite{18}. As a result, the renormalized energies are defined by
\begin{equation}
E_{0,\mathrm{ren}}^{(1/2)}=-2\hbar\left[\sum_{n=1}^{\infty}n(n+1)\omega_{n}^{(1/2)}-
\int_{0}^{\infty}t(t+1)\omega_{t}^{(1/2)}dt\right],
\label{0.6}
\end{equation}
\begin{eqnarray}
E_{0,\mathrm{ren}}^{(1)}=\hbar\left[\sum_{n=1}^{\infty}(n^2-1)\omega_{n}^{(1)}-
\int_{0}^{\infty}(t^2-1)\omega_{t}^{(1)}dt\right].
\label{0.7}
\end{eqnarray}
The differences between the sums and the integrals can be simply evaluated by using the
following Abel-Plana formulas \cite{03,05,37,38}:
\begin{eqnarray}
\sum_{n=1}^{\infty}\Phi\left(n\right)-\int_{0}^{\infty}\Phi\left(t\right)dt=
-\frac{1}{2}\Phi\left(0\right)+i\int_{0}^{\infty}\frac{\Phi\left(it\right)-
\Phi\left(-it\right)}{e^{2\pi t}-1}dt,
\label{0.8}
\end{eqnarray}
\begin{equation}
\sum_{n=1}^{\infty}\Phi\left(n+\frac{1}{2}\right)-\int_{0}^{\infty}\Phi\left(t\right)dt=
-i\int_{0}^{\infty}\frac{\Phi\left(it\right)-\Phi\left(-it\right)}{e^{2\pi t}+1}dt.
\label{0.9}
\end{equation}
Equations (\ref{0.8}) and (\ref{0.9}) are convenient in the cases of boson and fermion fields,
respectively. The Abel-Plana formulas (\ref{0.8}) and (\ref{0.9}) are valid for functions
that
are analytic in the right-half plane including the imaginary axis.

The application of Eq. (\ref{0.9}) to (\ref{0.6}) with
\begin{eqnarray}
\Phi\left(n+\frac{1}{2}\right)&=&\Phi_{1}^{(1/2)}\left(n+\frac{1}{2}\right)=n(n+1)
\left(n+\frac{1}{2}\right) \nonumber \\
&=&\left(n+\frac{1}{2}\right)^{3}-\frac{1}{4}\left(n+\frac{1}{2}\right)
\label{1.0}
\end{eqnarray}
leads to \cite{18,31}
\begin{eqnarray}
E_{0,\mathrm{ren}}^{(1/2)}=\frac{17\hbar c}{480 a}.
\label{1.1}
\end{eqnarray}
In a similar way, using Eq. (\ref{0.8}) in (\ref{0.7}) with
\begin{eqnarray}
\Phi\left(n\right)=\Phi_{1}^{(1)}\left(n\right)=n(n^2-1)=n^3-n,
\label{1.2}
\end{eqnarray}
one obtains \cite{18,31}
\begin{eqnarray}
E_{0,\mathrm{ren}}^{(1)}=\frac{11\hbar c}{120 a}.
\label{1.3}
\end{eqnarray}
As a result, the total free energy for the neutrino and electromagnetic fields in Einstein cosmological
model is given by
\begin{eqnarray}
F_{\mathrm{tot}}^{(s)}(T)=E_{0,\mathrm{ren}}^{(s)}+\Delta F^{(s)}(T),
\label{1.4}
\end{eqnarray}
where all terms on the right-hand side are defined in Eqs. (\ref{0.3}), (\ref{0.5}), (\ref{1.1}) and (\ref{1.3}).

Another important characteristic of the equilibrium state of the field at nonzero tempera\-ture is
the internal energy which is connected with the expectation value of the stress-energy tensor and
the free energy in the following way \cite{39,40}:
\begin{eqnarray}
U_{\mathrm{tot}}^{(s)}(T)=V\left\langle T_{0}^{(s)0}\right\rangle_{\mathrm{tot}}=
-T^2\frac{\partial}{\partial T}\left[\frac{F_{\mathrm{tot}}^{(s)}(T)}{T}\right].
\label{1.5}
\end{eqnarray}
Using Eqs. (\ref{0.3}), (\ref{0.5}), (\ref{1.4}) and (\ref{1.5}), we obtain for the internal energy
of the neutrino and electromagnetic fields
\begin{eqnarray}
U_{\mathrm{tot}}^{(s)}(T)=E_{0,\mathrm{ren}}^{(s)}+\Delta U^{(s)}(T),
\label{1.6}
\end{eqnarray}
where
\begin{eqnarray}
\Delta U^{(1/2)}(T)=4\hbar\sum_{n=1}^{\infty}
\frac{n(n+1)\omega_{n}^{(1/2)}}{e^{(\hbar\omega_{n}^{(1/2)}/k_B T)}+1},
\label{1.7}
\end{eqnarray}
\begin{eqnarray}
\Delta U^{(1)}(T)=2\hbar\sum_{n=1}^{\infty}\frac{(n^2-1)\omega_{n}^{(1)}}{e^{(\hbar\omega_{n}^{(1)}/k_B T)}-1}.
\label{1.8}
\end{eqnarray}

Now we proceed with the definitions of the Casimir free energy and internal energy at nonzero temperature.
These definitions go back to the Lifshitz theory \cite{41}, where the Casimir free energy of the
fluctuating electromagnetic field between two semispaces separated by a gap of width $a$ is obtained
by subtracting the free energy of an unbounded Minkowski space. In other words, not only an (infinite)
renormalization of the zero-point energy is done, but the thermal correction $\Delta F^{(s)}(T)$ undergoes
finite renormalization by subtracting the contribution of the black-body radiation \cite{05}.
This contribution is proportional to the volume of the gap between semispaces. The same definition for
the thermal Casimir effect was used in thermal quantum field theory for the configuration of two parallel
ideal metal plates \cite{05,41}. What is more, for configurations with a finite volume, such as ideal metal
rectangular boxes, it was shown \cite{42} that to obtain the Casimir free energy one should subtract from
$\Delta F^{(s)}(T)$ two more terms of quantum nature proportional to the surface area of the box and to
the perimeter of its edges. This is in line with Ref. \cite{43}, which demonstrated that in the
asymptotic limit of high temperature the total free energy in any restricted volume contains the
following three types of terms depending on the Planck constant:
\begin{eqnarray}
\alpha_{0}^{(s)}\frac{(k_B T)^4}{(\hbar c)^3}, \;\;\;\;\;\;\;\;
\alpha_{1}^{(s)}\frac{(k_B T)^3}{(\hbar c)^2}, \;\;\;\;\;\;\;\; \alpha_{2}^{(s)}\frac{(k_B T)^2}{\hbar c},
\label{1.9}
\end{eqnarray}
where the coefficients $\alpha_{0}^{(s)}$, $\alpha_{1}^{(s)}$, $\alpha_{2}^{(s)}$ depend on the spin
of the field and are expressed in terms of the heat kernel coefficients \cite{05}. Some of these
coefficients may be equal to zero. Specifically, for a conformal massless scalar field in the Einstein
cosmological model it is true that \cite{30}
\begin{eqnarray}
\alpha_{0}^{(0)}=-\frac{\pi^2}{90}V, \;\;\;\;\;\;\;\; \alpha_{1}^{(0)}=0, \;\;\;\;\;\;\;\; \alpha_{2}^{(0)}=0.
\label{2.0}
\end{eqnarray}

It can be seen that in our case the terms (\ref{1.9}) in the free energy are contained in the integral
\begin{eqnarray}
\int_{0}^{\infty}\Phi_{2}^{(s)}(t)dt,
\label{2.1}
\end{eqnarray}
where the function $\Phi_{2}^{(s)}(t)$ for the neutrino and electromagnetic fields in accordance
 with Eqs. (\ref{0.3}) and (\ref{0.5}) is given by
\begin{eqnarray}
\Phi_{2}^{(1/2)}(t)=-4k_{B}T\left(t^2-\frac{1}{4}\right)\mathrm{ln}\left[1+e^{-(\hbar ct/ak_BT)}\right],
\label{2.2}
\end{eqnarray}
\begin{eqnarray}
\Phi_{2}^{(1)}(t)=2k_{B}T(t^2-1)\mathrm{ln}\left[1-e^{-(\hbar ct/ak_BT)}\right].
\label{2.3}
\end{eqnarray}
Expanding the logarithms in power series and integrating, one obtains \cite{44}
\begin{eqnarray}
\int_{0}^{\infty}\Phi_{2}^{(1/2)}(t)dt=-\frac{7\pi^4 a^3}{90}\frac{(k_BT)^4}{(\hbar c)^3}+
\frac{\pi^2 a}{12}\frac{(k_BT)^2}{\hbar c},
\label{2.4}
\end{eqnarray}
\begin{eqnarray}
\int_{0}^{\infty}\Phi_{2}^{(1)}(t)dt=-\frac{2\pi^4 a^3}{45}\frac{(k_BT)^4}{(\hbar c)^3}+
\frac{\pi^2 a}{3}\frac{(k_BT)^2}{\hbar c}.
\label{2.5}
\end{eqnarray}

As a result, the Casimir free energy is defined as
\begin{eqnarray}
F_{C}^{(s)}(T)=E_{0,\mathrm{ren}}^{(s)}+\Delta F_{C}^{(s)}(T),
\label{2.6}
\end{eqnarray}
where the Casimir thermal correction is given by
\begin{eqnarray}
\Delta F_{C}^{(s)}(T)=\Delta F^{(s)}(T)-\int_{0}^{\infty}\Phi_{2}^{(s)}(t)dt.
\label{2.7}
\end{eqnarray}
The definition (\ref{2.7}) generalizes the approach previously followed in the Lifshitz theory and
in thermal quantum field theory for ideal metal plates and rectangular boxes. As can be seen in
Eq. (\ref{2.4}) and (\ref{2.5}), for neutrino and electromagnetic fields in the Einstein model we have
$$\alpha_{0}^{(1/2)}=-\frac{7\pi^2}{180}V, \;\;\;\;\;\;\;\; \alpha_{1}^{(1/2)}=0, \;\;\;\;\;\;\;\;
 \alpha_{2}^{(1/2)}=\frac{\pi^2a}{12}, \nonumber $$
\begin{equation}
\alpha_{0}^{(1)}=-\frac{\pi^2}{45}V, \;\;\;\;\;\;\;\; \alpha_{1}^{(1)}=0, \;\;\;\;\;\;\;\;
\alpha_{2}^{(1)}=\frac{\pi^2a}{3}.
\label{2.8}
\end{equation}
In this manner, the definition of the thermal correction in the Casimir free energy presumes the
subtraction of not only the contribution of the black-body radiation in free Minkowski space, as
it is true for the scalar field [see Eq. (\ref{2.0})], but one more term of quantum nature which
is present in the total free energy.
In Secs. III and IV it is shown that the asymptotic expressions of the total free energy
$F_{\mathrm{tot}}^{(s)}$ at high $T$ do not contain any more power-type terms of quantum nature.

The Casimir contribution to the internal energy (and respective Casimir contribution to the
thermal stress-energy tensor) can be defined in a similar way. It is given by
\begin{eqnarray}
U_{C}^{(s)}(T)=E_{0,\mathrm{ren}}^{(s)}+\Delta U_{C}^{(s)}(T),
\label{2.9}
\end{eqnarray}
where
\begin{eqnarray}
\Delta U_{C}^{(s)}(T)=\Delta U^{(s)}(T)-\int_{0}^{\infty}\Phi_{3}^{(s)}(t)dt.
\label{3.0}
\end{eqnarray}
In accordance with Eqs. (\ref{1.7}) and (\ref{1.8}), the function $\Phi_{3}^{(s)}(t)$ for the
neutrino and electromagnetic fields is given by
\begin{eqnarray}
\Phi_{3}^{(1/2)}(t)=\frac{\hbar c}{a}\;\frac{t(4t^2-1)}{e^{(\hbar ct/ak_BT)}+1},
\label{3.1}
\end{eqnarray}
\begin{eqnarray}
\Phi_{3}^{(1)}(t)=\frac{2\hbar c}{a}\;\frac{t(t^2-1)}{e^{(\hbar ct/ak_BT)}-1}.
\label{3.2}
\end{eqnarray}
The integrals subtracted in Eq. (\ref{3.0}) are simply calculated as
\begin{eqnarray}
\int_{0}^{\infty}\Phi_{3}^{(1/2)}(t)dt=\frac{7\pi^4 a^3}{30}\frac{(k_BT)^4}{(\hbar c)^3}-
\frac{\pi^2 a}{12}\frac{(k_BT)^2}{\hbar c},
\label{3.3}
\end{eqnarray}
\begin{eqnarray}
\int_{0}^{\infty}\Phi_{3}^{(1)}(t)dt=\frac{2\pi^4 a^3}{15}\frac{(k_BT)^4}{(\hbar c)^3}-
\frac{\pi^2 a}{3}\frac{(k_BT)^2}{\hbar c}.
\label{3.4}
\end{eqnarray}
It is easily seen that the integrals (\ref{2.4}) and (\ref{2.5}) subtracted from the total free
energy and the respective integrals (\ref{3.2}) and (\ref{3.3}) subtracted from the total internal
energy satisfy the same Eq. (\ref{1.5}) as the total quantities.

In the end of this section, we consider other components of the stress-energy tensor diffe\-rent
from $00$-component defined in Eq. (\ref{1.5}). In the spatially homogeneous isotropic metrics of
the space-time under consideration the stress-energy tensor is diagonal and has equal spatial components
\begin{eqnarray}
P_{\mathrm{tot}}^{(s)}(T)=-\left\langle T_{i}^{(s)i}\right\rangle_{\mathrm{tot}},
\label{3.5}
\end{eqnarray}
where $P_{\mathrm{tot}}^{(s)}(T)$ is the pressure. Both the total and Casimir pressures are
expressed using the respective free energies as
\begin{equation}
P_{\mathrm{tot}}^{(s)}(T)=-\frac{\partial F_{\mathrm{tot}}^{(s)}(T)}{\partial V},\;\;\;\;\;\;\;\;
P_{C}^{(s)}(T)=-\frac{\partial F_{C}^{(s)}(T)}{\partial V}.
\label{3.6}
\end{equation}

Considering Eqs. (\ref{0.3}), (\ref{0.5}), (\ref{1.1}), (\ref{1.3}), and (\ref{1.4}),
we obtain for the total pressure defined in Eqs. (\ref{3.6}), the following results
\begin{equation}
P_{\mathrm{tot}}^{(1/2)}(T)=\frac{17 \hbar c}{2880\pi^2 a^4}+\frac{2 \hbar}{3\pi^2 a^3}
\sum_{n=1}^{\infty}\frac{n(n+1)\omega_{n}^{(1/2)}}{e^{(\hbar\omega_{n}^{(1/2)}/k_BT)}+1},
\label{3.7}
\end{equation}
\begin{eqnarray}
P_{\mathrm{tot}}^{(1)}(T)=\frac{11 \hbar c}{720\pi^2 a^4}+\frac{\hbar}{3\pi^2 a^3}
\sum_{n=1}^{\infty}\frac{(n^2-1)\omega_{n}^{(1)}}{e^{(\hbar\omega_{n}^{(1)}/k_BT)}-1}.
\label{3.8}
\end{eqnarray}
Comparing Eq. (\ref{3.7}) with Eqs. (\ref{1.5})--(\ref{1.7}) and Eq. (\ref{3.8}) with
 Eqs. (\ref{1.5}), (\ref{1.6}) and (\ref{1.8}), we get the equation of state
\begin{eqnarray}
P_{\mathrm{tot}}^{(s)}(T)=\frac{1}{3}\varepsilon_{\mathrm{tot}}^{(s)}(T),
\label{3.9}
\end{eqnarray}
where the total energy density is just given by
\begin{eqnarray}
\varepsilon_{\mathrm{tot}}^{(s)}(T)=\left\langle T_{0}^{(s)0}\right\rangle_{\mathrm{tot}}=\frac{U^{(s)}(T)}{V},
\label{4.0}
\end{eqnarray}
in accordance with Eq. (\ref{1.5}). Using Eqs. (\ref{2.4}), (\ref{2.5}), (\ref{3.3}) and (\ref{3.4}),
it can be easily verified that the same equation of state is satisfied for the Casimir quantities
defined in Eqs. (\ref{2.6}), (\ref{2.7}), (\ref{2.9}), (\ref{3.0}) and (\ref{3.6}):
\begin{eqnarray}
P_{C}^{(s)}(T)=\frac{1}{3}\varepsilon_{C}^{(s)}(T).
\label{4.1}
\end{eqnarray}
Here, the Casimir energy density is the $00$-component of the diagonal Casimir stress-energy tensor
\begin{eqnarray}
\varepsilon_{C}^{(s)}(T)=\left\langle T_{0}^{(s)0}\right\rangle_{C}\equiv\frac{U_{C}^{(s)}(T)}{V}.
\label{4.2}
\end{eqnarray}
The other components of this tensor are given by $-P_{C}^{(s)}(T)$.

\section{Calculation of the total and Casimir free energy and internal energy for neutrino field}

Here, we calculate the free energy and internal energy for the neutrino field in Einstein and Friedmann
cosmological models. We begin from the representation of the Casimir free energy $F_{C}^{(1/2)}(T)$
as given by Eq. (\ref{2.6}) with $s=1/2$, where the energy at zero temperature is expressed by
Eq. (\ref{1.1}) and the Casimir thermal correction by Eqs. (\ref{0.3}), (\ref{2.2}) and (\ref{2.7}).
It is convenient to calculate the difference between the sum in Eq. (\ref{0.3}) and the integral in
 Eq. (\ref{2.7}) by using the Abel-Plana formula (\ref{0.9}) with $\Phi(t)=\Phi_{2}^{(1/2)}(t)$
 defined in Eq. (\ref{2.2}). This function is analytic in the right-half plane including the
 imaginary frequency axis. Expanding the logarithm in power series, one obtains
\begin{equation}
\Phi_{2}^{(1/2)}(t)=-4k_BT\left(t^2-\frac{1}{4}\right)
\sum_{n=1}^{\infty}\frac{(-1)^{n+1}}{n}e^{-(\hbar cnt/ak_BT)}
\label{4.3}
\end{equation}
and
\begin{equation}
\Phi_{2}^{(1/2)}(it)-\Phi_{2}^{(1/2)}(-it)=-8ik_BT\left(t^2+\frac{1}{4}\right)
\sum_{n=1}^{\infty}\frac{(-1)^{n+1}}{n}\sin\frac{\hbar cnt}{ak_BT}.
\label{4.4}
\end{equation}
Then, calculating the integral on the right-hand side of Eq. (\ref{0.9}), we arrive at the
representation of the Casimir free energy in terms of the double sum:
\begin{eqnarray}
&F_{C}^{(1/2)}(T)&=\frac{17\hbar c}{480a}+\frac{16\hbar c}{a}\sum_{n=1}^{\infty}(-1)^{n+1}
\sum_{m=1}^{\infty}(-1)^{m+1}\left\{\frac{(\hbar cn/ak_BT)^2-3(2\pi m)^2}{[(\hbar cn/ak_BT)^2+
(2\pi m)^2]^3}\right.\nonumber\\
& &\left. -\frac{1}{8}\;\frac{1}{(\hbar cn/ak_BT)^2+(2\pi m)^2}\right\}.
\label{4.5}
\end{eqnarray}

In Eq. (\ref{4.5}) one can carry out the summation either in $m$ or in $n$. The resulting
representations for the Casimir free energy are, strictly speaking, equivalent, but convenient
for obtaining the asymptotic limits in the case of low and high temperature, respectively.
The summation in $m$ can be performed by using the formula
\begin{eqnarray}
& &\sum_{m=1}^{\infty}(-1)^{m+1}\left[\frac{x^2-3y^2m^2}{(x^2+y^2m^2)^3}-\frac{1}{8(x^2+y^2m^2)}\right]
\nonumber\\
& &=\frac{1}{8x^4}\left\{4-\pi^3x^3[3+\cosh(2\pi x/y)]\mathrm{csch}^3(\pi x/y)\right\}\nonumber\\
& &-\frac{y-\pi x\mathrm{csch}(\pi x/y)}{16x^2y},
\label{4.6}
\end{eqnarray}
where $x=\hbar cn/ak_BT$ and $y=2\pi$. By performing the summation in $m$ in Eq. (\ref{4.5}) with
the help of Eq. (\ref{4.6}), one arrives at
\begin{eqnarray}
F_{C}^{(1/2)}(T)&=&\frac{17\hbar c}{480a}+\frac{7\pi^4a^3}{90}\frac{(k_BT)^4}{(\hbar c)^3}-
\frac{\pi^2a}{12}\frac{(k_BT)^2}{\hbar c}\nonumber\\
&-&k_BT\sum_{n=1}^{\infty}\frac{(-1)^{n+1}}{n}\frac{1}{\sinh^3(\hbar cn/2ak_BT)}.
\label{4.7}
\end{eqnarray}
From this equation, in the low temperature limit, where $\hbar c/ak_BT\rightarrow\infty$, we have
\begin{eqnarray}
F_{C}^{(1/2)}(T)&=&\frac{17\hbar c}{480a}+\frac{7\pi^4a^3}{90}\frac{(k_BT)^4}{(\hbar c)^3}-
\frac{\pi^2a}{12}\frac{(k_BT)^2}{\hbar c}\nonumber\\
&-&8k_BTe^{-(3\hbar c/2ak_BT)}.
\label{4.8}
\end{eqnarray}
This result is similar to the low-temperature behavior of the Casimir free energy in the case of two
parallel ideal metal plates \cite{05,41}.

From Eq. (\ref{4.7}) it is easy to obtain the exact expression for the total free energy.
Using Eqs. (\ref{1.4}), (\ref{2.4}) and (\ref{2.7}) one has all power-type contributions in $T$
canceled, and then the total free energy is given by
\begin{equation}
F_{\mathrm{tot}}^{(1/2)}(T)=\frac{17 \hbar c}{480a}-k_BT\sum_{n=1}^{\infty}\frac{(-1)^{n+1}}{n}
\frac{1}{\sinh^3(\hbar cn/2ak_BT)}.
\label{4.9}
\end{equation}
Thus, the thermal correction in the total free energy at low temperature is exponentially small:
\begin{equation}
F_{\mathrm{tot}}^{(1/2)}(T)=\frac{17 \hbar c}{480a}-8k_BTe^{-(3\hbar c/2ak_BT)}.
\label{5.0}
\end{equation}

To obtain the asymptotic behavior of the Casimir and total free energy at high temperature, we perform
the summation in $n$ in Eq. (\ref{4.5}) first. This leads to the following representation for the
Casimir free energy:
\begin{eqnarray}
& &F_{C}^{(1/2)}(T)=\frac{17 \hbar c}{480a}+\frac{a^2}{16\pi^4}\frac{(k_BT)^3}{(\hbar c)^2}
\sum_{m=1}^{\infty}\frac{(-1)^{m+1}}{m^4\sinh^3(2\pi^2mak_BT/\hbar c)}\nonumber\\
& &\times\left\{4\pi^2m\left[4\pi^4m^2+\frac{(\hbar c)^2}{(ak_BT)^2}(2+\pi^2m^2)\right]
\cosh\left(\frac{4\pi^2mak_BT}{\hbar c}\right)\right. \nonumber\\
& &\left. + 3\frac{(\hbar c)^3}{(ak_BT)^3}(6+\pi^2m^2)\sinh\left(\frac{2\pi^2mak_BT}{\hbar c}
\right)\right.\nonumber\\
& &\left. +4\pi^2m\left[12m^2\pi^4-\frac{(\hbar c)^2}{(ak_BT)^2}(2+\pi^2m^2)+4\pi^2
\frac{\hbar c}{ak_BT}m\sinh\left(\frac{4\pi^2mak_BT}{\hbar c}\right)\right]\right.\nonumber\\
& &\left. -\frac{(\hbar c)^3}{(ak_BT)^2}(6+\pi^2m^2)\sinh\left(\frac{6\pi^2mak_BT}{\hbar c}\right)\right\}.
\label{5.1}
\end{eqnarray}
The same expression can be equivalently rewritten in the form
\begin{eqnarray}
& &F_{C}^{(1/2)}(T)=\frac{17 \hbar c}{480a}+\frac{a^2}{4\pi^4}\frac{(k_BT)^3}{(\hbar c)^2}
\sum_{m=1}^{\infty}\frac{(-1)^{m+1}}{m^4}\nonumber\\
& &\times\left\{-\frac{(\hbar c)^3}{(ak_BT)^3}(6+\pi^2m^2)+2m\pi^2\mathrm{csch}
\left(\frac{2\pi^2mak_BT}{\hbar c}\right)\right. \nonumber\\
& &\times\left. \left[\pi^2\frac{\hbar cm}{ak_BT}\left(\frac{\hbar cm}{ak_BT}+4
\mathrm{coth}\left(\frac{2\pi^2mak_BT}{\hbar c}\right)\right)+2\left(\frac{(\hbar c)^2}{(ak_BT)^2}+
2m^2\pi^4\right.\right.\right.\nonumber\\
& &\left.\left.\left. 4m^2\pi^4\mathrm{csch}^2\left(\frac{2\pi^2mak_BT}{\hbar c}\right)\right)\right]\right\}.
\label{5.2}
\end{eqnarray}

From Eqs. (\ref{5.1}) or (\ref{5.2}) in the limit of high temperature one obtains
\begin{equation}
F_{C}(T)=4\pi^2a^2\frac{(k_BT)^3}{(\hbar c)^2}e^{-(2\pi^2ak_BT/\hbar c)}.
\label{5.3}
\end{equation}
Note that in the limit of high $T$ the zero-temperature contribution to the Casimir free energy was
canceled by the corresponding term in the thermal correction. We emphasize that the Casimir free
energy at high $T$ is exponentially small. It does not contain the classical term proportional to
$k_BT$ (see Section IV), as is expected for a spinor field.

The total free energy is obtained by adding Eq. (\ref{2.4}) to Eqs. (\ref{5.1}) or (\ref{5.2}).
The expressions obtained for $F_{\mathrm{tot}}^{(1/2)}(T)$ are equivalent to the more compact
 Eq. (\ref{4.9}). The asymptotic behavior of the total free energy at high $T$ is obtained by
 adding Eqs. (\ref{2.4}) and (\ref{5.3}).

A quantity closely related to the free energy is the entropy. The total and the Casimir entropies
for both the neutrino and electromagnetic fields are defined as
\begin{equation}
S_{\mathrm{tot}}^{(s)}(T)=-\frac{\partial{F_{\mathrm{tot}}^{(s)}(T)}}{\partial T}, \;\;\;\ S_{C}^{(s)}(T)=
-\frac{\partial{F_{C}^{(s)}(T)}}{\partial T}.
\label{5.4}
\end{equation}
The exact expression for the Casimir entropy of the neutrino field can be obtained from
Eq. (\ref{4.7}). It is given by
\begin{eqnarray}
& &S_{C}^{(1/2)}(T)=\pi^2k_B\frac{ak_BT}{3\hbar c}\left[\frac{1}{2}-
\frac{14\pi^2}{15}\frac{(ak_BT)^2}{(\hbar c)^2}\right]\nonumber\\
& &+k_B\sum_{n=1}^{\infty}\frac{(-1)^{n}}{n}\left[\mathrm{csch}^3(\hbar cn/2ak_BT)+
\frac{3\hbar cn}{2ak_BT}\cosh\left(\frac{\hbar cn}{2ak_BT}\right)\right.\nonumber\\
& &\times\left.\mathrm{csch}^4(\hbar cn/2ak_BT)\right].
\label{5.5}
\end{eqnarray}
When the temperature vanishes, the asymptotic expression for the Casimir entropy is given by
\begin{eqnarray}
S_{C}^{(1/2)}(T)&=&\pi^2k_B\frac{ak_BT}{3\hbar c}\left[\frac{1}{2}-\frac{14\pi^2}{15}
\frac{(ak_BT)^2}{(\hbar c)^2}\right]\nonumber\\
&+&\frac{12\hbar c}{aT}e^{-(3\hbar c/2ak_BT)}.
\label{5.6}
\end{eqnarray}
As is seen from Eq. (\ref{5.6}), the Casimir entropy goes to zero when $T$ goes to zero, i.e.,
the third law of thermodynamics (the Nernst heat theorem) is satisfied \cite{39,40}. The total
entropy is obtained from Eqs. (\ref{4.9}) and (\ref{5.4}). It is given by Eq. (\ref{5.5}) with
the first term on the right-hand side omitted. The asymptotic behavior of the total entropy at
low $T$ is given by the exponentially small term in Eq. (\ref{5.6}). Thus, the Nernst heat theorem
also holds for the total entropy of neutrino field.

Now we turn to the consideration of the internal energy. The total internal energy for the neutrino
field is given by Eqs. (\ref{1.6}) and (\ref{1.7}). The Casimir internal energy can be calculated
 by Eqs. (\ref{2.9}) and (\ref{3.0}) with the function $\Phi_{3}^{(1/2)}(t)$ defined in
 Eq. (\ref{3.1}). The calculation follows the same steps as shown above for the free energy and
 uses the Abel-Plana formula (\ref{0.9}). The same results can also be obtained from the respective
 expressions for the free energy by using Eq. (\ref{1.5}) and similar equations for the Casimir
 quantities. Thus, from Eqs. (\ref{4.9}) and (\ref{1.5}), we find the following result for the
 total internal energy:
\begin{equation}
U_{\mathrm{tot}}^{(1/2)}(T)=\frac{17\hbar c}{480a}+\frac{3\hbar c}{2a}\sum_{n=1}^{\infty}(-1)^{n+1}
\frac{\cosh(\hbar cn/2ak_BT)}{\sinh^4(\hbar cn/2ak_BT)}.
\label{5.7}
\end{equation}
In the limit of low temperature, from Eq. (\ref{5.7}), one obtains
\begin{equation}
U_{\mathrm{tot}}^{(1/2)}(T)=\frac{17\hbar c}{480a}+\frac{12\hbar c}{a}e^{-(3\hbar c/2ak_BT)}.
\label{5.8}
\end{equation}
In the limit of high temperature the asymptotic behavior of the total internal energy is
\begin{eqnarray}
&U_{\mathrm{tot}}^{(1/2)}(T)&=\frac{7\pi^4a^3}{30}\frac{(k_BT)^4}{(\hbar c)^3}-\frac{\pi^2a}{12}
\frac{(k_BT)^2}{\hbar c}\nonumber\\
& &+8\pi^4a^3\frac{(k_BT)^4}{(\hbar c)^3}e^{-(2\pi^2ak_BT/\hbar c)}.
\label{5.9}
\end{eqnarray}
Dividing both sides of Eqs. (\ref{5.8}) and (\ref{5.9}) by the spatial volume $V$, we get
\begin{equation}
\varepsilon_{\mathrm{tot}}^{(1/2)}(T)=\frac{17\hbar c}{960\pi^2a^4}+\frac{6\hbar c}{\pi^2a^4}
e^{-(3\hbar c/2ak_BT)},
\label{6.0}
\end{equation}
\begin{eqnarray}
&\varepsilon_{\mathrm{tot}}^{(1/2)}(T)&=\frac{7\pi^2}{60}\frac{(k_BT)^4}{(\hbar c)^3}-
\frac{1}{24a^2}\frac{(k_BT)^2}{\hbar c}\nonumber\\
& &+4\pi^2\frac{(k_BT)^4}{(\hbar c)^3}e^{-(2\pi^2ak_BT/\hbar c)}
\label{6.1}
\end{eqnarray}
at low and high temperature, respectively. These coincide with the asymptotic behavior of the energy
density for the neutrino field found in Ref. \cite{31} (note that the exponentially small terms were
omitted in \cite{31}).

The exact expression for the Casimir internal energy is obtained by subtracting Eq. (\ref{3.3})
from Eq. (\ref{5.7}). As a result, at low and high temperature we have, respectively,
\begin{eqnarray}
&U_{C}^{(1/2)}(T)&=\frac{17\hbar c}{480a}-\frac{7\pi^4a^3}{30}\frac{(k_BT)^4}{(\hbar c)^3}+
\frac{\pi^2a}{12}\frac{(k_BT)^2}{\hbar c}\nonumber\\
& &+\frac{12\hbar c}{a}e^{-(3\hbar c/2ak_BT)},
\label{6.2}
\end{eqnarray}
\begin{eqnarray}
U_{C}^{(1/2)}(T)=8\pi^4a^3\frac{(k_BT)^4}{(\hbar c)^3}e^{-(2\pi^2ak_BT/\hbar c)}.
\label{6.3}
\end{eqnarray}
Thus, for the neutrino field the Casimir internal energy at high $T$ is exponentially small.
The limiting values for the Casimir energy density are obtained from Eqs. (\ref{6.2}) and (\ref{6.3})
after the division by $V$. As to the total and Casimir pressures, they can be obtained from the
 equations of state (\ref{3.9}) and (\ref{4.1}).

\section{Calculation of the free energy and internal energy for electromagnetic field}
\hspace{-0.7cm} The case of the electromagnetic field is more complicated because the function
$\Phi_{2}^{(1)}(t)$ defined in Eq. (\ref{2.3}) goes to infinity when $t$ goes to zero. This prevents
direct application of the Abel-Plana formula (\ref{0.8}) for the calculation of the Casimir free
 energy of the electromagnetic field defined in Eqs. (\ref{2.6}) and (\ref{2.7}). A similar problem
 arises in the calculation of the total and Casimir internal energies given by Eqs. (\ref{1.6}),
 (\ref{1.8}) and (\ref{2.9}), (\ref{3.0}), respectively. The point is that the function
 $\Phi_{3}^{(1)}(t)$ in Eq. (\ref{3.2}), determining the thermal correction (\ref{1.8}) in the
 internal energy, has poles along the imaginary frequency axis. This also prevents the application
 of the Abel-Plana formula in its simplest form (\ref{0.8}).

We start from the total free energy (\ref{0.1}), (\ref{0.5}), and expand the logarithm into the
power series to obtain
\begin{equation}
F_{\textrm{tot}}^{(1)}(T)=\frac{11\hbar c}{120a}-2k_BT\sum_{m=1}^{\infty}\frac{1}{m}
\sum_{n=1}^{\infty}(n^2-1)e^{-(\hbar cnm/ak_BT)}.
\label{6.4}
\end{equation}
Here, the sum with respect to $n$ can be calculated leading to
\begin{equation}
F_{\textrm{tot}}^{(1)}(T)=\frac{11\hbar c}{120a}-2k_BT\sum_{m=1}^{\infty}
\frac{3e^{\hbar cm/ak_BT}-1}{m(e^{\hbar cm/ak_BT}-1)^3}.
\label{6.5}
\end{equation}
In accordance with Eq. (\ref{2.7}) the Casimir free energy for the electromagnetic field is obtained
 subtraction of Eq. (\ref{2.5}) from Eq. (\ref{6.5}):
\begin{eqnarray}
&F_{C}^{(1)}(T)&=\frac{11\hbar c}{120a}+\frac{2\pi^4a^3}{45}\frac{(k_BT)^4}{(\hbar c)^3}-
\frac{\pi^2a}{3}\frac{(k_BT)^2}{\hbar c}\nonumber\\
& &-2k_BT\sum_{m=1}^{\infty}\frac{3e^{\hbar cm/ak_BT}-1}{m(e^{\hbar cm/ak_BT}-1)^3}.
\label{6.6}
\end{eqnarray}
Equations (\ref{6.5}) and (\ref{6.6}) are convenient for obtaining the asymptotic expressions at
low temperature. For the total free energy one obtains from Eq. (\ref{6.5})
\begin{equation}
F_{\textrm{tot}}^{(1)}(T)=\frac{11\hbar c}{120a}-6k_BTe^{-(2\hbar c/ak_BT)}.
\label{6.7}
\end{equation}
It is seen that total correction to the energy at zero temperature is exponentially small.
The low temperature behavior of the Casimir free energy is given by the difference between
Eqs. (\ref{6.7}) and (\ref{2.5}). In this case, there are power-type corrections to the energy
at zero temperature.

Before considering the asymptotic behavior of the free energy at high $T$, we obtain the exact
expression for the internal energy and its asymptotic behaviors at low and high temperatures.
Using the expansion in power series in Eq. (\ref{1.8}), we can rewrite Eq. (\ref{1.6}) in the form
\begin{equation}
U_{\textrm{tot}}^{(1)}(T)=\frac{11\hbar c}{120a}+\frac{2\hbar c}{a}\sum_{m=1}^{\infty}
\sum_{n=1}^{\infty}n(n^2-1)e^{-(\hbar cmn/ak_BT)}.
\label{6.8}
\end{equation}
By performing the summation in $n$ we obtain the following exact expression:
\begin{equation}
U_{\textrm{tot}}^{(1)}(T)=\frac{11\hbar c}{120a}+\frac{12\hbar c}{a}\sum_{m=1}^{\infty}
\frac{e^{2\hbar cm/ak_BT}}{(e^{\hbar cm/ak_BT}-1)^4}.
\label{6.9}
\end{equation}
The Casimir internal energy is given by Eqs. (\ref{2.9}) and (\ref{3.0}). Taking into account
Eq. (\ref{3.4}) we have
\begin{eqnarray}
&U_{C}^{(1)}(T)&=\frac{11\hbar c}{120a}-\frac{2\pi^4a^3}{15}\frac{(k_BT)^4}{(\hbar c)^3}+
\frac{\pi^2a}{3}\frac{(k_BT)^2}{\hbar c}\nonumber\\
& &+\frac{12\hbar c}{a}\sum_{m=1}^{\infty}\frac{e^{2\hbar cm/ak_BT}}{(e^{\hbar cm/ak_BT}-1)^4}.
\label{7.0}
\end{eqnarray}
Equations (\ref{6.9}) and (\ref{7.0}) are convenient for obtaining the asymptotic expressions
at low $T$. Thus, from Eq. (\ref{6.9}) it follows
\begin{equation}
U_{\textrm{tot}}^{(1)}(T)=\frac{11\hbar c}{120a}+\frac{12\hbar c}{a}e^{-(2\hbar c/ak_BT)}.
\label{7.1}
\end{equation}
After subtraction of Eq. (\ref{3.4}) from Eq. (\ref{7.1}), the asymptotic expression for the Casimir
internal energy at low $T$ is obtained. All these expressions for the internal energy are connected
 with respective expressions for the free energy, obtained above, by Eq. (\ref{1.5}) and a similar
 equation for the Casimir quantities.

Now we turn to the Casimir internal energy in the form of Eqs. (\ref{2.9}) and (\ref{3.0}), and
calculate it in a different way, which allows us to obtain the high temperature limit. The poles
of the function $\Phi_{3}^{(1)}(t)$ defined in Eq. (\ref{3.2}) are located at the points
$t=it_{l}=2\pi ilak_BT/\hbar c$, where $l=\pm 1, \pm 2,...$ . In addition, at $t=0$ this
function takes a nonzero value $\Phi_{3}^{(1)}(0)=-2k_BT$. The presence of the poles along the
imaginary frequency axis makes the standard Abel-Plana formula inapplicable. There is, however,
a generalization of this formula for the case when there are poles on the imaginary frequency axis.
It reads \cite{37}
\begin{eqnarray}
\sum_{n=1}^{\infty}&\Phi(n)&-\int_{0}^{\infty}\Phi(t)dt=-\frac{1}{2}\Phi(0)-
\pi\sum_{l}\mathop{\mathrm{Res}}_{t=it_{l}}\left[\frac{\Phi(t)e^{i\pi t}}{\sin(\pi t)}\right]\nonumber\\
& & + i\int_{0}^{\infty}\frac{\Phi(it)-\Phi(-it)}{e^{2\pi t}-1}dt.
\label{7.2}
\end{eqnarray}
Here, it is assumed that summation is done only over the positive $l$ and the function $\Phi(t)$
satisfies the condition
\begin{equation}
\Phi(t)=\Phi(-t)+o\left[\frac{1}{(t-it_{l})}\right] \;\;\;\ \textrm{when} \;\;\;\ t\rightarrow it_{l}.
\label{7.3}
\end{equation}

We apply Eq. (\ref{7.2}) to the function $\Phi(t)=\Phi_{3}^{(1)}(t)$ defined in Eq. (\ref{3.2}).
In doing so, we take into account that
\begin{eqnarray}
&-&\pi\sum_{l=1}^{\infty}\mathop{\mathrm{Res}}_{t=it_{l}}
\left[\frac{\Phi_{3}^{(1)}(t)e^{i\pi t}}{\sin(\pi t)}\right]\nonumber\\
&=&\frac{32\pi^4a^3(k_BT)^4}{(\hbar c)^3}\sum_{l=1}^{\infty}\frac{l^3}{e^{4\pi^2lak_BT/\hbar c}-1}\nonumber\\
&+&\frac{8\pi^2a(k_BT)^2}{\hbar c}\sum_{l=1}^{\infty}\frac{l}{e^{4\pi^2lak_BT/\hbar c}-1}
\label{7.4}
\end{eqnarray}
and
\begin{equation}
\Phi_{3}^{(1)}(it)-\Phi_{3}^{(1)}(-it)=\frac{2\hbar c}{a}it(t^2+1).
\label{7.5}
\end{equation}
Calculating the integral on the right-hand side of Eq. (\ref{7.2}),
\begin{equation}
i\int_{0}^{\infty}\frac{\Phi_{3}^{(1)}(it)-\Phi_{3}^{(1)}(-it)}{e^{2\pi t}-1}dt=-\frac{11}{120}\frac{\hbar c}{a},
\label{7.6}
\end{equation}
we find that it cancels the contributions of the Casimir energy at zero temperature
$E_{0,\textrm{ren}}^{(1)}$ defined in Eq. (\ref{1.3}). Then from Eqs. (\ref{2.9}) and (\ref{3.0}) one obtains
\begin{eqnarray}
&U_{C}^{(1)}(T)&=k_BT+\frac{32\pi^4a^3(k_BT)^4}{(\hbar c)^3}
\sum_{l=1}^{\infty}\frac{l^3}{e^{4\pi^2lak_BT/\hbar c}-1}\nonumber\\
& &+\frac{8\pi^2a(k_BT)^2}{\hbar c}\sum_{l=1}^{\infty}\frac{l}{e^{4\pi^2lak_BT/\hbar c}-1}.
\label{7.7}
\end{eqnarray}
This is an alternative exact expression for the Casimir internal energy. Numerical computations show
that is gives the same values of $U_{\textrm{Cas}}^{(1)}(T)$ as Eq. (\ref{7.0}) over the entire range
of temperatures. The total internal energy $U_{\textrm{tot}}^{(1)}(T)$ is obtained by adding
Eq. (\ref{3.4}) to Eq. (\ref{7.7}).

Equation (\ref{7.7}) is convenient for obtaining the asymptotic limit at high temperature.
Thus, from this equation the asymptotic expression for the Casimir internal energy at high $T$ is given by
\begin{equation}
U_{C}^{(1)}(T)=k_BT+\frac{32\pi^4a^3(k_BT)^4}{(\hbar c)^3}e^{-(4\pi^2ak_BT/\hbar c)}.
\label{7.8}
\end{equation}
The asymptotic expression at high $T$ for the total internal energy is obtained from Eq. (\ref{7.8})
 by adding Eq. (\ref{3.4}):
\begin{eqnarray}
&U_{\textrm{tot}}^{(1)}(T)&=\frac{2\pi^4 a^3}{15}\frac{(k_BT)^4}{(\hbar c)^3}-
\frac{\pi^2 a}{3}\frac{(k_BT)^2}{\hbar c}+k_BT \nonumber\\
& &+\frac{32\pi^4a^3(k_BT)^4}{(\hbar c)^3}e^{-(4\pi^2ak_BT/\hbar c)}.
\label{7.9}
\end{eqnarray}
From Eq. (\ref{7.8}) it is seen that the electromagnetic Casimir internal energy at high temperature
 has a classical limit \cite{33}, which does not depend on $\hbar$ and $c$. The total internal
 energy (\ref{7.9}) also contains a classical term equal to $k_BT$. The respective asymptotic
 expression for the total internal energy density is
\begin{eqnarray}
&\varepsilon_{\textrm{tot}}^{(1)}(T)&=\frac{U_{\textrm{tot}}^{(1)}(T)}{V}=\frac{\pi^2}{15}
\frac{(k_BT)^4}{(\hbar c)^3}- \frac{1}{6a^2}\frac{(k_BT)^2}{\hbar c}+\frac{k_BT}{2\pi^2a^3} \nonumber\\
& &+16\pi^2\frac{(k_BT)^4}{(\hbar c)^3}e^{-(4\pi^2ak_BT/\hbar c)}.
\label{8.0}
\end{eqnarray}
Note that the high temperature behavior of $\varepsilon_{\textrm{tot}}^{(1)}$ obtained in Ref. \cite{31}
does not contain the classical term. This is explained by the omission of the term
$\Phi_{3}^{(1)}(0)/2=-k_BT$ in the Poisson summation formula (Eq. (4) of Ref. \cite{31}), which
is equivalent to the Abel-Plana formula used here.

The obtained expression (\ref{7.8}) for the Casimir internal energy at high $T$ can be used for
the determination of the high temperature behavior of the Casimir free energy, which is still unknown.
Using the thermodynamic connection (\ref{1.5}) between the free energy and internal energy, we have
\begin{equation}
F_{C}^{(1)}(T)=-T\left[\int \frac{U_{C}^{(1)}(T)}{T^2}dT+k_BR\right],
\label{8.1}
\end{equation}
where $R$ is an arbitrary dimensionless constant independent of $T$. Substituting Eq. (\ref{7.8}) in
 Eq. (\ref{8.1}) and keeping only the high order terms, one arrives at
\begin{eqnarray}
&F_{C}^{(1)}(T)&=-k_BT\mathrm{ln}\frac{ak_BT}{\hbar c}-Rk_BT\nonumber\\
& &+8\pi^2a^2\frac{(k_BT)^3}{(\hbar c)^2}e^{-(4\pi^2ak_BT/\hbar c)}.
\label{8.2}
\end{eqnarray}
The value of the constant $R$ can be determined from the exact expression for $F_{C}^{(1)}(T)$
[see Eq. (\ref{6.6})]. Computations show that the exact expression (\ref{6.6}) leads to the same
values of the Casimir free energy as the asymptotic expression (\ref{8.2}) up to six significant
figures for the values of parameter $ak_BT/\hbar c\geq 1$ when $R=1.77698$. 
Thus, the Casimir free
energy of the electromagnetic field at low $T$ contains not only the classical (entropic) term,
 but the logarithmic contribution as well. The same holds for the Casimir free energy inside an
 ideal metal spherical shell \cite{05, 45, 46} (at zero temperature the Casimir energy
for fields of different spins in spherically symmetric cavities was considered in
Refs.~\cite{47,48}). 
The low temperature behavior of the total free
 energy is obtained from Eq. (\ref{8.2}) by adding Eq. (\ref{2.5}).

In the end of this section we consider the total and Casimir entropy for the 
electromagnetic field
in Einstein and closed Friedmann cosmological model. At low temperature from 
Eqs. (\ref{5.4}) and
(\ref{6.7}) one obtains the following main contribution to the total entropy:
\begin{equation}
S_{\mathrm{tot}}^{(1)}(T)=\frac{12\hbar c}{aT}e^{-(2\hbar c/ak_BT)}.
\label{8.3}
\end{equation}
For the Casimir entropy at low $T$ using Eq. (\ref{2.5}) we arrive at
\begin{eqnarray}
&S_{C}^{(1)}(T)&=2\pi^2k_B\frac{ak_BT}{3\hbar c}\left[1-\frac{4\pi^2}{15}
\frac{(ak_BT)^2}{(\hbar c)^2}\right]\nonumber\\
& &+\frac{12\hbar c}{aT}e^{-(2\hbar c/ak_BT)}.
\label{8.4}
\end{eqnarray}
As is seen from Eqs. (\ref{8.3}) and (\ref{8.4}), both the total and Casimir entropies go to zero
when the temperature vanishes, in accordance with the Nernst heat theorem.

Using the above results, the expressions for the total and Casimir pressures of electromagnetic
field and their behaviors at low and high temperature can be obtained from Eqs. (\ref{3.9}) and (\ref{4.0}).

\section{Conclusions and discussion}

In the foregoing we have investigated the thermal Casimir effect for the neutrino and electromagnetic
fields in the Einstein cosmological model. The results obtained are also valid in the closed Friedmann
cosmological model where they should be considered as complementary to 
the terms describing the vacuum
 polarization and particle creation caused by the nonstationary regime 
of the metric. For both fields
 under consideration we found general expressions for the total 
internal energy, energy density and
  pressure (considered earlier) and for the total free energy. 
In all cases, we separated the Casimir
  contributions from the obtained quantities by means of an additional subtraction procedure.

Both the total and Casimir internal energy and free energy were represented 
in the form of single sums.
The asymptotic expressions for these sums were found in the limiting cases 
of low and high temperature.
For the neutrino field, our results for the total internal energy (energy density) 
are in agreement
 with those obtained in Ref. \cite{31}. We have also found the exponentially 
small corrections to the
 results of Ref. \cite{31}. Our results for the Casimir free energy of neutrino 
field are similar to
 those obtained for configurations with material boundaries. 
Specifically, we have shown that the
 Casimir free energy of neutrino field does not possess a classical 
limit at high temperature and
 the Casimir entropy satisfies the Nernst heat theorem.

For the electromagnetic field, our calculation result for the total internal energy differs from
the result of Ref. \cite{31} by a classical term which arises from 
the contribution of zero arguments
in the Poisson and Abel-Plana summation formulas. We have shown 
that for the electromagnetic field at
 high temperature the total internal energy includes not only the 
Planck-type terms, as it was believed
  previously, but also the linear in temperature classical term. 
For the Casimir free energy of the
  electromagnetic field in Einstein cosmological model we have proved 
that there are both classical
  and logarithmic in temperature terms in the limit of high temperature, 
as it holds for the thermal
  Casimir effect inside an ideal metal sphere. Both the total and Casimir 
entropies were demonstrated
  to be in agreement with the Nernst heat theorem.

In future, it would be interesting to determine the influence of the revealed classical 
term on the
cosmological evolution, where the total internal energy plays the role of a source, and consider
 multi-dimensional generalizations of the obtained results for application in 
brane-world scenarios.
Specifically, it seems advantageous to generalize 
the results of Refs.~\cite{23,24} to the case of nonzero
temperature.

\section*{Acknowledgments}

V.B.B., V.M.M. and C.R. were supported by CNPq (Brazil). H.F.M. was supported
by CAPES (Brazil). The authors are grateful to G. L. Klimchitskaya and 
A. A. Saharian for numerous
helpful discussions. V.M.M. is also grateful to Federal University of 
Paraíba, where this work was done,
for kind hospitality.

\end{document}